\documentclass[]{interact}

\usepackage{epstopdf}
\usepackage[caption=false]{subfig}

\usepackage[numbers,sort&compress]{natbib}
\bibpunct[, ]{[}{]}{,}{n}{,}{,}
\renewcommand\bibfont{\fontsize{10}{12}\selectfont}
\makeatletter
\def\NAT@def@citea{\def\@citea{\NAT@separator}}
\makeatother
\usepackage[numbers,sort&compress]{natbib}
\bibpunct[, ]{[}{]}{,}{n}{,}{,}
\renewcommand\bibfont{\fontsize{10}{12}\selectfont}
\makeatletter
\def\NAT@def@citea{\def\@citea{\NAT@separator}}
\makeatother

\theoremstyle{plain}

\theoremstyle{definition}

\usepackage[left,pagewise]{lineno}

\theoremstyle{remark}

\usepackage[normalem]{ulem}
\usepackage{graphicx}%
\usepackage{multirow}%
\usepackage{amsmath,amssymb,amsfonts}%
\usepackage{amsthm}%
\usepackage{mathrsfs}%
\usepackage[title]{appendix}%
\usepackage{xcolor}%
\usepackage{textcomp}%
\usepackage{manyfoot}%
\usepackage{booktabs}%
\usepackage{algorithm}%
\usepackage{algorithmicx}%
\usepackage{algpseudocode}%
\usepackage{listings}%
\usepackage{placeins}
\usepackage{tabularx}
\usepackage{array}
\usepackage{xcolor}
\usepackage{soul}
\usepackage{bm} 
\usepackage{gensymb}
\usepackage[colorlinks=true, allcolors=blue]{hyperref}

\begin{document}

\articletype{Original Report}

\title{Intrinsic grain-size gradients upon grain growth near a free surface}

\author{
\name{Jing Tang\textsuperscript{a}\thanks{Corresponding Author. Email: jing.tang@unileoben.ac.at}, Runlu Yan\textsuperscript{a}, Dong-lan Zhang\textsuperscript{b}, Ronald Schnitzer\textsuperscript{a}, Lorenz Romaner\textsuperscript{a}, Marlene Kapp\textsuperscript{c}, Marco Salvalaglio\textsuperscript{d,e}, Oliver Renk\textsuperscript{a}}
\affil{\textsuperscript{a} Department of Materials Science, Montanuniversität Leoben, Roseggerstraße 12, 8700, Leoben, Austria.}
\affil{\textsuperscript{b} Jiangxi Province Key Laboratory of Additive Manufacturing of Implantable Medical Device, Jiangxi University of Science and Technology, 330013, Nanchang, China.}
\affil{\textsuperscript{c} Erich-Schmid-Institute of Materials Science, Austrian Academy of Sciences, 8700, Leoben, Austria.}
\affil{\textsuperscript{d} Institute of Scientific Computing, TU Dresden, 01062, Dresden, Germany.}
\affil{\textsuperscript{e} Dresden Center for Computational Materials Science (DCMS), TU Dresden, 01062, Dresden, Germany.}
}

\maketitle
\thispagestyle{plain}

\begin{abstract}

Grain growth in polycrystals is generally attributed to mean curvature flow, although recent studies have suggested that elastic effects associated with shear-coupled grain boundary migration may also influence the kinetics. Here, we characterize the effect of free surfaces on grain growth kinetics in high-purity bulk polycrystalline nickel. By analyzing the microstructural evolution on cross sections of 1 mm thick annealed specimens from the surface to the interior, as well as through in-plane investigations on specimens with varying thickness (1 mm, 40 $\mu$m, and 10 $\mu$m), an intrinsic grain-size gradient was identified, characterized by a gradual increase in grain size towards the interior. Interestingly, this grading was not restricted to the very surface but continued to depths more than five layers of grains, where effects from thermal grooves are considered negligible. We demonstrate that this behavior is significantly affected by elastic relaxation at the free surface, which alters the internal stress fields generated by shear-coupled grain boundary migration. These findings emphasize the relevance of free surfaces to the microstructural evolution of polycrystals.
\end{abstract}


\section{Introduction}
Microstructures of metallic materials evolve at high temperature, largely through the migration of interfaces between crystalline domains with different orientations, namely grain boundaries (GBs). This ubiquitous process leads to an increase in average grain size (grain growth), which is key to material processing and performance. The resulting microstructure largely dictates not only the mechanical but also functional properties of engineering materials \cite{hall1954variation,petch1953cleavage,cheng2013grain,armstrong1970influence,wei2021grain,li1989effect,herzer2013modern}. Therefore, understanding the mechanisms underlying grain growth and related phenomena, such as recrystallization, is key to predict process-microstructure-property relationships
\cite{hillert1965theory, najafkhani2021recent, srolovitz1984computer,holm2010grain,burke1952recrystallization}.

GBs are extended defects possessing excess energy. Their migration is driven by the reduction of interfacial energy. It has long been described by an equation of motion corresponding to mean-curvature flow, i.e. $v = M \kappa \gamma$. Here, $v$ is the GB migration velocity, with $\kappa $ and $\gamma$ being the mean local curvature and GB energy, respectively. $M$ is the GB mobility; it is a temperature-dependent proportionality constant depending on the temperature and activation energy of grain boundary diffusion via an Arrhenius-type equation \cite{rollett2021grain,humphreys2012recrystallization},  and accounting for diffusive exchange of atoms or atomic units across the GB \cite{burke1952recrystallization,gottstein2009grain,rollett2021grain,humphreys2012recrystallization}. In recent years, increasing evidence has shown that a description of GB migration based solely on the reduction of interface energy (capillarity) does not capture the complexity of microstructural evolution in polycrystalline solids. Results using emerging X-ray–based techniques \cite{bernier2020high, oddershede2022advanced} to non-destructively map the microstructural evolution in 4D (i.e., three spatial dimensions plus time or temperature), allowed to analyze about 50,000 GB segments of polycrystalline nickel \cite{bhattacharya2021grain}. It was found that GB velocity is weakly, if not at all, correlated with curvature. Similar results were obtained for other materials, such as iron \cite{xu2024grain} and strontium titanate \cite{muralikrishnan2023observations}. Detailed analyses of GB motion further indicate that even flat GBs can migrate rather fast, or curved segments could also move away from their center of curvature \cite{muralikrishnan2023observations}, eluding comparisons with microstructure evolution simulations, even when carefully accounting for experimental morphological features \cite{peng2022comparison}.

Several key aspects have been identified as essential for properly describing grain growth in polycrystals. GB anisotropy, arising from the atomistic structure of grain boundaries (GBs), strongly influences GB dynamics \cite{rohrer2011grain,rohrer2023grain} and is crucial for capturing morphological features. In addition, GB migration has been shown to be coupled to shear deformation of grains (shear-coupling) \cite{cahn2006coupling,khater2012disconnection,han2018grain}, which leads to the development of internal stresses \cite{thomas2017reconciling} that have also been measured during microstructural evolution, for instance in recrystallized grains \cite{zhang2022local,lindkvist20233d,yildirim20253d}.
A unified description of these effects can be achieved by recognizing that GB migration is mediated by the flow of disconnections \cite{ashby1972boundary,hirth1973grain,han2018grain}. These defects can be described as steps with dislocation character constrained to the GB plane. Their coupling to stress, via the Peach–Koehler force analogous to that acting on lattice dislocations, accounts for mechanically induced grain growth in nanocrystalline materials even at cryogenic temperatures \cite{rajabzadeh2013elementary,kapp2020plastic,frazer2020cryogenic}.
Moreover, thermally driven grain growth is also associated with stress buildup through the shear-coupling mechanism, which significantly affects microstructural evolution. This has been demonstrated in recent simulation and experimental studies \cite{thomas2017reconciling,qiu2025grain,tian2024grain,tang2026atomic} and represents a fundamental mechanism for understanding the phenomenology of grain growth. Examples include the weak correlation between GB velocity and curvature \cite{bhattacharya2021grain, qiu2025grain}, grain rotation during annealing observed in nanocrystalline platinum thin films \cite{tian2024grain}, and discrete surface height changes associated with GB migration, measured for instance on nickel surfaces with grain sizes on the order of 10 $\mu$m \cite{tang2026atomic}.

Grain growth is typically characterized under the assumption of bulk behavior. However, free surfaces are inherently present and can significantly influence the microstructural evolution. A well-known effect is the formation of thermal grooves at specimen surfaces \cite{mullins1958effect}, which tend to retard grain growth. The impact of free surfaces has been extensively studied in thin polycrystalline systems \cite{kaplan2013review,verma2021grain,verma2023effect}, whereas their impact on microstructures of bulk specimens remains less explored. This has actually practical impact for very thin products, e.g., wires and thin sheets. The recent advances in the understanding of grain growth outlined above further suggest that additional mechanisms must be considered, particularly the role of internal stresses. 

In this work, we investigate the influence of free surfaces on grain growth in high-purity polycrystalline nickel samples with varying thicknesses. We reveal the emergence of intrinsic grain-size gradients associated with the presence of a free surface, characterized by a progressive slowdown of grain growth towards the specimen surface.To discuss and understand these observations, we employ a continuum modeling framework that incorporates the fundamental mechanisms of grain boundary migration. This approach highlights novel aspects of how free surfaces modify internal stress fields. We demonstrate experimentally that the influence of the free surface extends several grain layers into the bulk, at least five, well beyond the region directly affected by thermal grooving, which is confined to the first one or two layers at most. This deeper grain-size gradient is attributed to internal stress relaxation resulting from shear-coupled GB migration. Within a minimal theoretical framework, we show that variations in the relative orientation between the shear stress generated by GB migration and the free surface can either accelerate or retard grain growth. These effects extend significantly beyond those expected from purely capillarity-driven mechanisms.

\section{Materials and Methods}
\subsection{Experiments}

To study the impact of free surfaces on grain growth, we consider different pure nickel specimens. A rather fine starting grain size of about 10 $\mu$m was targeted, as previous grain growth investigations for this grain size range already provided evidence of shear-coupled migration up to at least 300 \degree C \cite{tang2026atomic}.

Therefore, severely plastically deformed high-purity nickel (99.99\% from Good Fellow, UK) specimens served as the starting material for the grain growth study. A nickel disc with 30 mm diameter and about 8 mm height was subjected to quasi-constrained high-pressure torsion (HPT) at room temperature for 15 rotations. The corresponding strains of $\sim$100 allow for a fairly homogeneous ultra-fined grained structure ($\sim$200 nm grain size) throughout the disc, except for the very center part. The homogeneous ultra-fine grained structure consists predominantly of random high-angle grain boundaries and a weak deformation texture. This enables uniform growth behavior, starting already at moderate annealing temperatures above 150 \degree C \cite{renk2023anneal}. Growth continues up to 400 \degree C, the temperature range in which shear-coupled grain growth was observed \cite{tang2026atomic}, providing a suitable starting material for our study. Subsequently, specimens were cut from the HPT disc at radii $>$ 8 mm, to exclude effects from the less deformed sample center.

\begin{figure}[htp!]
    \centering
    \includegraphics[width=0.95\textwidth]{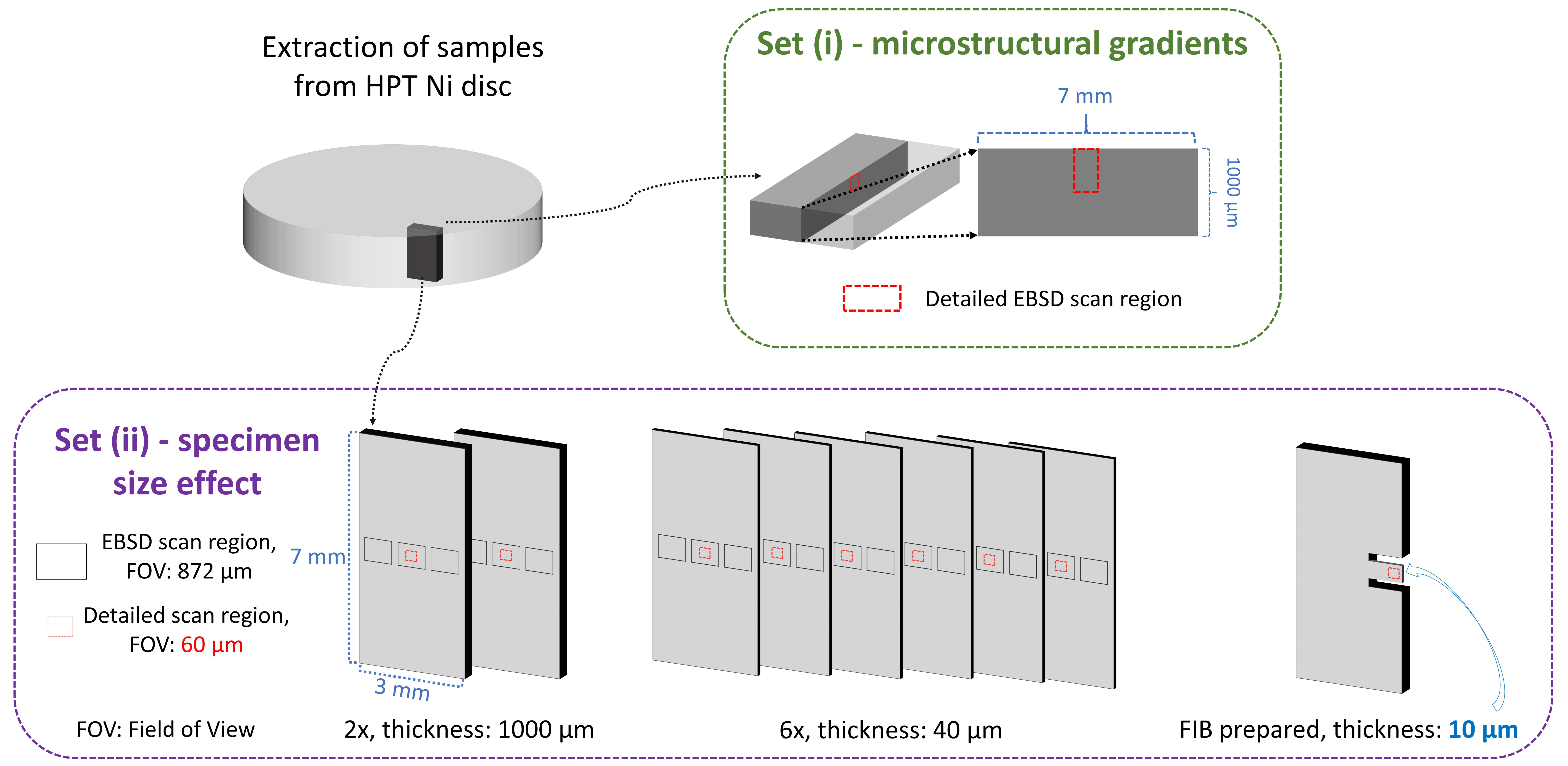}
    \caption{Schematic illustration of the sample extraction from the HPT disc and the two different set of specimens investigated in this work. Dimensions are not to scale.}
    \label{figure1}
\end{figure}

Two different sets of samples, each of them having a cross section of $3 \times 7$ mm$^{2}$ (with the long axis corresponding to the thickness of the original HPT disc), were extracted from the processed disc (see schematics in Fig.\ref{figure1}) and subjected to annealing at 400 \degree C in vacuum (HTM Reetz furnace, Germany) for 1~h. 
The two sample sets investigated here consisted of:

(i) 1 mm thick slice cut at mid-width after the heat treatment to analyze potential gradients in grain growth kinetics from the surface to the specimen interior within a single sample. Microstructures were investigated perpendicular to this cut at different depths from the surface, starting at the very surface to the center of the specimen (500 $\mu$m depth); see Fig.~\ref{figure1}.

(ii) multiple specimens with different thicknesses ranging from 1000 $\mu$m, over $\sim$40 $\mu$m, down to $\sim$ 10 $\mu$m thick specimens to analyze sample size effects on grain growth kinetics. Especially in the two thinner specimens, surface effects are expected to be more pronounced and, consequently, to exert a stronger influence on grain growth. To precisely control the thickness of the 40 $\mu$m thin specimens, a Struers Accustop was used. To account for potential volume effects six $\sim$40 $\mu$m thick specimens were prepared to cover similar volumes as for the 1 mm thick reference. The sample extraction and an overview of the two different sample sets are shown in a schematic in Fig.~\ref{figure1}.

For one of the $\sim$40 $\mu$m thick specimens a cross section was prepared after the heat treatment using a broad Ar-ion beam (ArBlade 5000 ion milling system, Hitachi, Japan) to measure the thickness and to determine the number of grains across the thickness. The two ultrathin nickel lamellae ($\sim$10 $\mu$m thickness) were prepared by femtosecond laser ablation (3D-Micromac microPREP PRO FEMTO), followed by focused ion beam (FIB) polishing using a FEI Versa 3D DualBeam microscope.

\begin{figure}[htp!]
    \centering
    \includegraphics[width=0.95\textwidth]{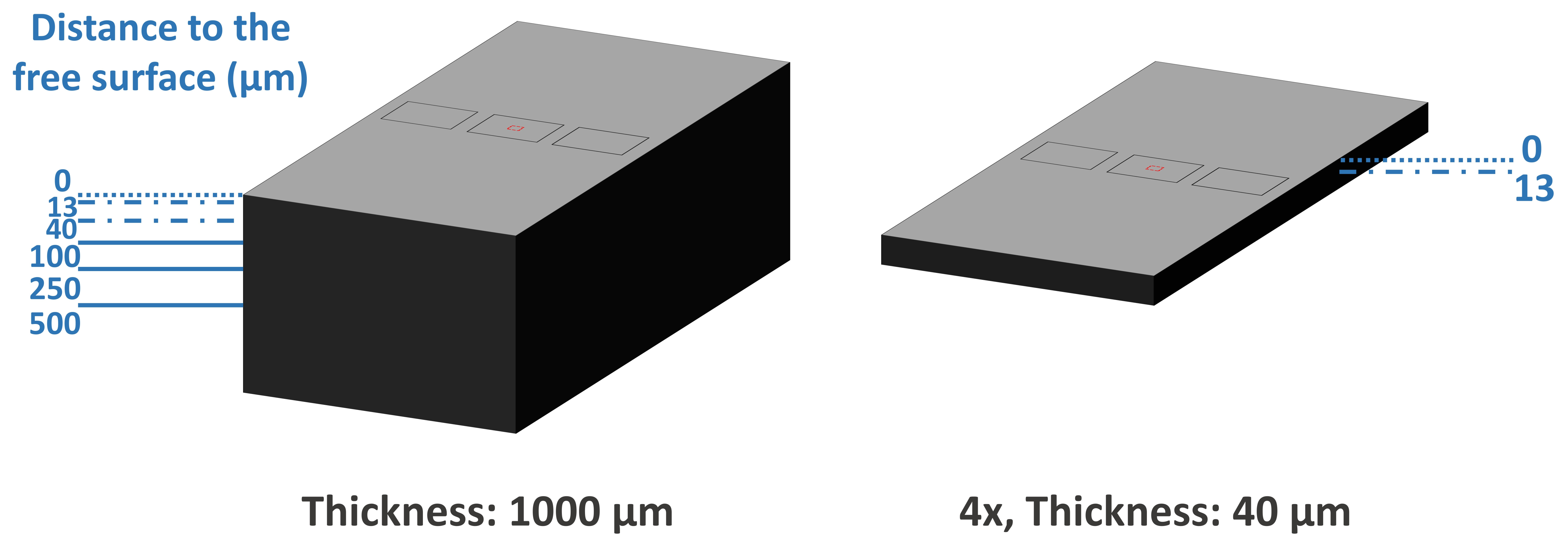}
    \caption{To separate grooving effects, and assess the microstructural evolution at different depths from the surface, specimens were sequentially characterized using EBSD. Layers were removed by electropolishing (thin specimens,40 $\mu$m) to depth of 13 $\mu$m and a combination of grinding and electropolishing in case of thick specimens (1000 $\mu$m) to depth of 0, 13, 40, 100, 250, 500 $\mu$m. Dimensions are not to scale.}
    \label{figure2}
\end{figure}

All specimens were ground and vibratory polished (VibroMet 2, Buehler, Germany) with OPU solution for 4~h, or electropolished (commercial A2 solution in Struers LectroPol-5 at a voltage of 20 V for 17 s with a flow rate of 12), to obtain a strain-free surface for subsequent EBSD analysis. Since the thin specimens had to be ground to the final thickness already before the annealing treatment, thermal grooving could affect the gathered EBSD data. Therefore, the thin specimens were either gently electro-polished or FIB-polished after annealing to remove the very surface layer, which was potentially affected by grooving. The removed thickness was determined by measuring the specimens/foils before and after electropolishing using a confocal laser scanning microscope (Keyence VK-X1100, Belgium). Two and four rounds of electropolishing with aforementioned parameters removed 13 $\mu$m and 40 $\mu$m thickness from the surface, respectively. Following each level of material removal, EBSD analysis was performed on the resultant fresh surface (i.e., after removal of 13 $\mu$mm and 40 $\mu$m, respectively). In addition to removing the effects of thermal grooving, this procedure enabled analysis of gradients in grain growth kinetics in the thin specimens, as shown schematically in Fig. 
\ref{figure2}. In case of the FIB-prepared lamellae, surface layers (about 4 $\mu$m in thickness on each side) were removed by additional FIB polishing after the heat treatment.

All microstructural investigations were based on EBSD data, acquired with a field-emission gun scanning electron microscope (Tescan Clara) equipped with an Oxford Symmetry S3 EBSD camera. The accelerating voltage and probe current for EBSD measurements were set to 20 kV and 10 nA using the ultrahigh resolution mode, and a suitable step size was carefully selected for individual EBSD scans (provided in the results section). For data analysis, the software package EDAX OIM Analysis 8 was used. A grain was defined as having a tolerance angle greater than 15 degrees and a minimum size of 5 pixels. Grains located at scan edges were included in the statistics \cite{zhu2020microstructure}. All investigations focused on the center of the specimens, i.e., at mid-height of the initial HPT disk.

\FloatBarrier

\subsection{Continuum modeling of grain boundary migration}
\label{sec:continuum-modeling}
To investigate the impact of a free surface on microstructure evolution, we employ a continuum model incorporating capillarity and shear-coupled grain boundary migration. We analyze scenarios capturing the essential surface effects, namely elastic relaxation due to the zero-traction boundary condition and surface-induced pinning \cite{mullins1958effect,frost1990simulation,barmak2006grain,verma2021grain}.

We describe an interface/GB $\Sigma$ through a smooth curve $\boldsymbol{\xi}(s)$ embedded in the $\hat{\mathbf{x}}-\hat{\mathbf{y}}$ plane with $s$ being a (scalar) parameter. As introduced in Ref.~\cite{han2022disconnection}, this description can represent the macroscopic limit of microscopic arrangements of \textit{disconnections} \cite{han2018grain}, each aligned with one of the low-energy GB orientations chosen as reference orientations $\mathbf{e}^{(k)}$ with $k=1,...,N$. For simplicity, we consider two perpendicular reference orientations ($N=2$), that approximate, e.g., [110] tilt GBs in FCC crystals (for an extension, see, e.g., \cite{qiu2023interface}). A GB with an arbitrary orientation can thus be described as a deviation from the nearest reference interfaces through a superposition of disconnections from the set of reference ones. 

The contribution of each reference is quantified by the densities $\rho^{(k)}$, determined geometrically from how many disconnections per unit length are required to transform the reference orientation into the actual orientation of $\Sigma$. These can be related to the tangent vector $\mathbf{l}$ of $\Sigma$ by
\begin{equation}
\mathbf{l}=  \left[ {\begin{array}{cc}
   \hat{\mathbf{x}} \cdot \hat{\mathbf{e}}^{(1)}  & \hat{\mathbf{x}} \cdot \hat{\mathbf{e}}^{(2)}\\
   \hat{\mathbf{y}} \cdot \hat{\mathbf{e}}^{(1)}  &  \hat{\mathbf{y}} \cdot  \hat{\mathbf{e}}^{(2)}\\
  \end{array} } \right]
  \left[ {\begin{array}{cc}
   -h^{(2)}\rho^{(2)}  \\
   h^{(1)}\rho^{(1)}  \\
  \end{array} } \right],
\end{equation}
with $h^{(k)}$ being the step height of the disconnections. 

Disconnections also possess a Burgers vector $\mathbf{b}^{(k)}$ that mediates shear deformation. Both $h^{(k)}\mathbf{e}^{(k)}$ and $\mathbf{b}^{(k)}$ are displacement shift complete (DSC) lattice vectors \cite{han2018grain}. Under the simplifying assumption of having one active disconnection mode per reference, namely the disconnection with the smallest $h^{(k)}$ and $|\mathbf{b}^{(k)}|$, and considering disconnection gliding along references only, the equation of motion (EOM) for $ \mathbf{p}\in \Sigma$ can be derived by considering the overdamped dynamics $\partial_t{\boldsymbol{\xi}}=-\mathbf{M} \cdot \delta E / \delta \boldsymbol{\xi}$ with $E$ comprising interface energy and the work done by gliding disconnections \cite{han2022disconnection}. It reads
\begin{equation}\label{eq:equation_of_motion}
    \partial_t{\mathbf{x}}=\mathbf{M} (\Gamma \kappa+ \boldsymbol{\tau} \cdot \boldsymbol{\beta}) \mathbf{\hat n},
\end{equation}
with parameters following from the corresponding values of the reference orientations, such as
$\mathbf{M}=(M^{(1)},M^{(2)})\mathbf{I}$ the mobility tensor, $\Gamma=\gamma(\phi)+\gamma''(\phi)$ the interface stiffness with $\gamma(\phi,\gamma^{(1)},\gamma^{(2)})$ the interface energy density and $\phi$ the GB inclination angle, $\hat{\mathbf{n}}$ the vector normal to $\Sigma$, $\kappa=\nabla_s \hat{\mathbf{n}}$ the curvature, $\boldsymbol{\tau}=(\tau^{(1)}, \tau^{(2)})$ the resolved shear stresses along the references, and $\boldsymbol{\beta}=(\beta^{(1)},\beta^{(2)})$ the shear coupling factors. The latter formally correspond to ${b}^{(k)}/h^{(k)}$ but could be considered as independent parameters \cite{qiu2025grain}.

The first term in Eq.~\eqref{eq:equation_of_motion} represents capillarity-driven contributions, while the second captures the coupling between interface motion and the stress field with \begin{equation}
    \tau^{(k)} = \frac{\sigma_{22} - \sigma_{11}}{2}\sin(2\phi^{(k)}) 
    + \sigma_{12}\cos(2\phi^{(k)}),
\end{equation}
the resolved shear stress (RSS) w.r.t reference orientations ($\phi^{(k)}$) and $\sigma_{ij}=\sigma_{ij}^{\rm ext}+\sigma_{ij}^{\rm int}$ the total stress including externally applied stresses ($\boldsymbol{\sigma}^{\rm ext}$) and internal stresses associated with the shear coupling mechanism ($\boldsymbol{\sigma}^{\rm int}$), mediated by disconnection flow along the GB. The latter, for an elastically isotropic system and assuming linear elasticity, can be evaluated efficiently by superimposing the elastic fields associated with the dislocation character of each disconnection in the system, which are known analytically \cite{han2022disconnection,sal2022disconnection}.

Here, we extend the framework to account for the presence of an infinitely extended free surface. In this setting, the analytic expressions for the elastic stress fields of dislocations are also known \cite{head1953edge,marzegalli2013onset}.
The internal (or self-) stress resulting from continuous densities of dislocations with Burgers vector $\mathbf{b}^{(k)}$ is given by
\begin{equation}\label{eq:stress}
    \sigma_{ij}^{\rm int}(s)
    = K \sum_{k=1}^{N} \int_{\Sigma} 
    \beta^{(k)}\, l_x \,\widetilde{\sigma}_{ij}\left(\boldsymbol{\xi}(s)-\boldsymbol{\xi}(s'),H(s'),\mathbf{b}^{(k)}\right)
    \, \mathrm{d}s',
\end{equation}
with $K=\mu/[2\pi(1-\nu)]$, $\mu$ the shear modulus, $\nu$ the Poisson ratio, $\widetilde{\boldsymbol{\sigma}}(x(s)-x(s'),H(s'))$ the elastic field in $x(s)$ generated by a dislocation in $x(s')$ at a distance $H(s')$ from a free surface with the normal $\hat{\boldsymbol{\nu}}$ parallel to the y-axis. It takes the form \cite{head1953edge,marzegalli2013onset}
\begin{equation}\label{eq:dislo_full}
\widetilde{\sigma}_{ij}(\mathbf{r},h,\mathbf{b}) =\sigma^{\mathrm{bulk}}_{ij}(\mathbf{r},\mathbf{b})- \sigma^{\mathrm{bulk}}_{ij}(\mathbf{r}-2h\hat{\mathbf{y}},\mathbf{b})
+ \sigma^{\mathrm{corr},\parallel}_{ij}(\mathbf{r},h,b_x)
+ \sigma^{\mathrm{corr},\perp}_{ij}(\mathbf{r},h,b_y),
\end{equation}
with $\sigma^{\mathrm{dislo}}$ being the bulk elastic field for a dislocation centered in the origin 
\begin{equation}
\begin{split}
\frac{1}{K}\sigma^{\rm bulk}_{xx}(\mathbf{r},\mathbf{b}) &= 
-b_x \frac{y\,(3x^2 + 3c^2 + y^2)}{r^4}+b_y\frac{ x\,(c^2 - y^2 + x^2)}{r^4},
\\
\frac{1}{K}\sigma^{\rm bulk}_{yy}(\mathbf{r},\mathbf{b}) &= 
-b_x\frac{y\,(c^2 - x^2 + y^2)}{r^4}+b_y\frac{x\,(3y^2 + 3c^2 + x^2)}{r^4},
\\
\frac{1}{K}\sigma^{\rm bulk}_{xy}(\mathbf{r},\mathbf{b}) &=b_x 
\frac{x\,(c^2 + x^2 - y^2)}{r^4}-b_y\frac{y\,(c^2 + y^2 - x^2)}{r^4},
\end{split}
\end{equation}
with $\mathbf{r}=(x,y)$, $\mathbf{b}=(b_x,b_y)$, $r=\sqrt{x^2+y^2+c^2}$, and $c$ a small parameter as introduced in \cite{cai2006non} to regularize the singularity at the core. $\sigma^{\rm bulk}$ enters the first two terms of Eq.\eqref{eq:dislo_full} as dislocation in the system and image dislocation, respectively \cite{head1953edge}.
The terms $\sigma^{\mathrm{corr},\parallel}_{ij}(\mathbf{x},h,b_x)$ and $\sigma^{\mathrm{corr},\perp}_{ij}(\mathbf{x},h,b_y)$
represents the correction required to fulfill the zero-traction boundary condition $\boldsymbol{\sigma}\cdot \hat{\boldsymbol{\nu}}=\mathbf{0}$ for an edge dislocation with Burgers vector parallel and perpendicular to the surface, respectively. They read
\begin{equation}
\begin{split}
\sigma^{{\rm corr},\parallel}_{xx}(\mathbf{r},h,b_x)
&=
-Kb_x\frac{2h}{r_h^6}
\bigl[
y (y-2h)^3
- 6 (y-h)(y-2h)x^2
+ x^4
\bigr], \\
\sigma^{{\rm corr},\parallel}_{yy}(\mathbf{r},h,b_x)
&=
Kb_x\frac{2h}{r_h^6}
\bigl[
(3y-4h)(y-2h)^3
- 6(y-h)(y-2h)x^2
- x^4
\bigr], \\
\sigma^{{\rm corr},\parallel}_{xy}(\mathbf{r},h,b_x)
&=
Kb_x\frac{4h (y-h)x}{r_h^6}
\bigl[3 (y-2h)^2 - x^2 \bigr], \\
\sigma^{{\rm corr},\perp}_{xx}(\mathbf{r},h,b_y)
&=
-Kb_y\frac{4xh}{r_h^6}
\bigl[
(y+h)(y-2h)^2
+ (-3y+5h)x^2
\bigr], \\
\sigma^{{\rm corr},\perp}_{yy}(\mathbf{r},h,b_y)
&=
Kb_y\frac{4xh (y-h)}{r_h^6}
\bigl[
3 (y-2h)^2 - x^2
\bigr], \\
\sigma^{{\rm corr},\perp}_{xy}(\mathbf{r},h,b_y)
&=
Kb_y\frac{2h}{r_h^6}
\bigl[
y(2h-y)^3
+ 6(-y+h)(2h-y)x^2
- x^4
\bigr].
\end{split}
\end{equation}
with $r_h=\sqrt{x^2+(y-2h)^2}$. The components of the stress field with and without a free surface are illustrated in Fig.~\ref{figure3}(a). We employ reduced units. All quantities with dimensions of length are scaled as  
$
\tilde{\mathbf{x}} = {\mathbf{x}}/{a}$ with $a$ the length of the DSC unit cell, $
\tilde{\kappa} = \kappa \alpha$,
$\Delta \tilde{t} = \Delta t\, M^{(1)} {\gamma^{(1)}}/{a^{2}}$, $\tilde{\gamma} = {\gamma^{(2)}}/{\gamma^{(1)}}$, $
\tilde{M} = {M^{(2)}}/{M^{(1)}}$, $
\tilde{\tau} = \tau\, {a}/{\gamma^{(1)}}$, $
\tilde{\psi} = \psi\, {a}/{\gamma^{(1)}}$. Also, we assume isotropic interface properties, i.e., ${\tilde{M}}=1$.

\subsection{Numerical simulations}
\label{sec:numerical-simulations}
\begin{figure}[htbp]
    \centering
\includegraphics[width=0.95\textwidth]{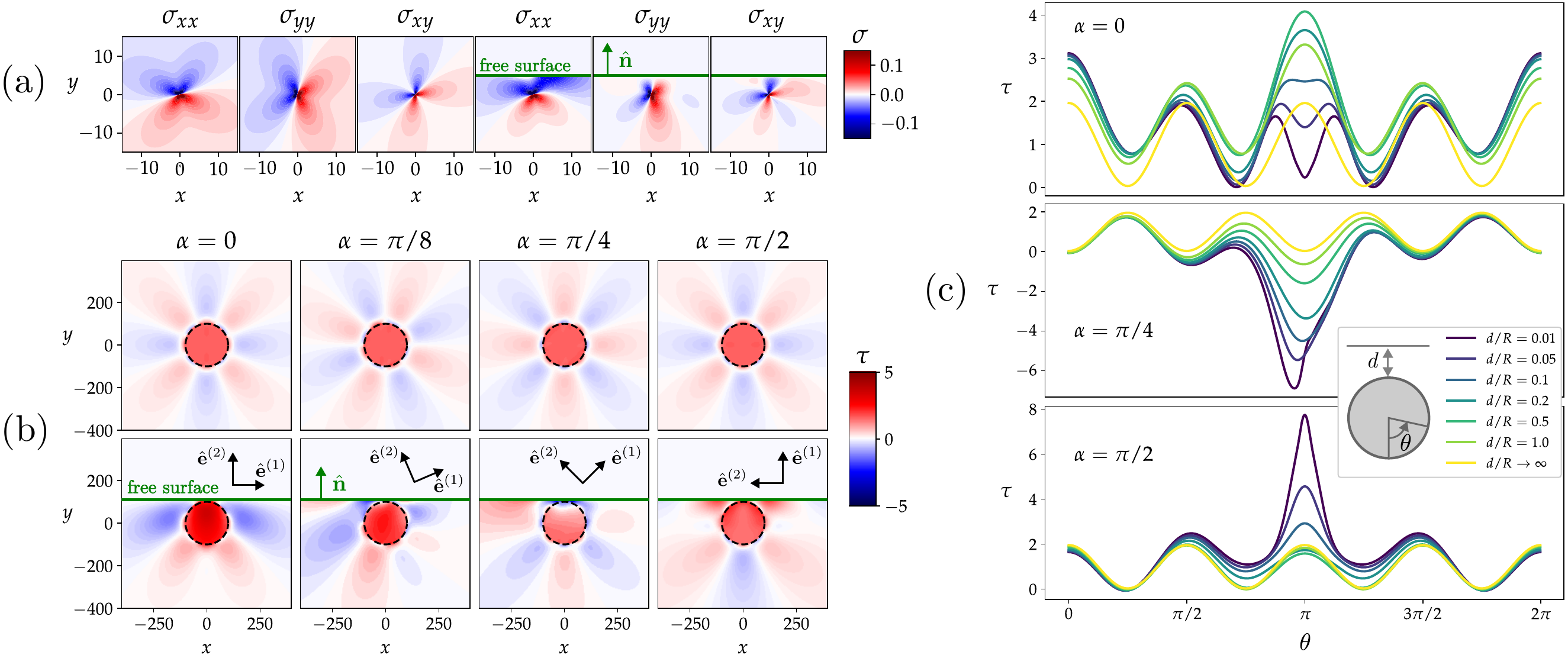}
    \caption{
    Impact of an infinite straight free surface on elastic fields. (a) Dislocation with $\mathbf{b}=(\sqrt{2}/2,\sqrt{2}/2)$ illustrated by stress field components with (three panels on the right) and without (three panels on the left) a free surface (green solid line) with $\hat{\mathbf{n}}=\hat{\mathbf{y}}$ at $y=5$. (b) Internal resolved shear stress (RSS) generated by disconnections (Eq.~\eqref{eq:stress}) over a circular grain boundary changing the orientation ($\alpha$) of references $\mathbf{e}^{(k)}$: $\mathbf{b}^{(1)}=1$, $\mathbf{b}^{(2)}=0$, $\mathbf{e}^{(1)}=[\cos(\alpha),\sin(\alpha)]$, $\mathbf{e}^{(1)}=[-\sin(\alpha),\cos(\alpha)]$. RSS is computed w.r.t reference $\mathbf{e}^{(1)}$. This setting mimics a grain with the same misorientation with respect to the surrounding matrix but different relative orientation ($\alpha$) with respect to the free surface. The fields with (bottom) and without (top) a free surface (with $\hat{\mathbf{n}}=\hat{\mathbf{y}}$) at $y=5$ are shown. (c) Resolved shear stress at the GB for three orientations $\alpha$. These are the terms $\tau^{(1)}\equiv\tau$ entering Eq.~\eqref{eq:equation_of_motion} for the chosen parameters.
    }
    \label{figure3}
\end{figure}

To examine how a free surface influences the internal stresses generated during shear-coupled GB migration, we consider an idealized configuration: a circular grain with radius $R=100$ is positioned at varying distances (\textit{d}) from, and with different orientations ($\alpha$) relative to the free surface. The resolved shear stress $\tau^{(1)}$ for some representative settings is illustrated in Fig.~\ref{figure3}(b). Note that rotating reference orientations leads to analogous (but rotated) stress fields in bulk, while the presence of the free surface with a fixed normal breaks such a rotational symmetry. In practice, varying the relative orientation of the shear deformation w.r.t. the surface normal leads to different elastic deformation of the crystal. The resolved shear stress at the GB entering the corresponding equation of motion is also shown in detail in Fig.~\ref{figure3}(c). Note that a non-trivial behavior emerges where, depending on the relative orientation of the induced shear and of the surface, the RSS may vary significantly and also change signs, involving both regions close to (at $\theta=\pi$) and far away from the free surface.

Numerical simulations tracking the morphological evolution of the grains beyond the initial conditions illustrated in Fig.~\ref{figure3}(b) and \ref{figure3}(c) are reported below in the Results section. They are obtained by exploiting a phase-field (PF) model (restricted to a single grain). In brief, we consider a phase field $\varphi(\mathbf{r})$ which is $1$ inside the grain, $0$ outside the grain, with a smooth transition in between. We solve the dynamic equation numerically
\begin{equation}\label{eq:pf-dynamics}
\frac{\partial \varphi}{\partial t}=\nabla^2 \varphi - \frac{1}{\varepsilon} H'(\varphi) + |\nabla \varphi| \Lambda \bar \tau 
\end{equation}
with $\varepsilon=10$ a (numerical) parameter controlling the thickness of the diffuse interface between grain and the matrix (one order of magnitude smaller than $R$), $H(\varphi)=18\varphi^2(1-\varphi)^2$, $\bar \tau$ the resolved shear stress computed for the $\varphi=0.5$ isoline (which approximates $\Sigma$) and extended along the normal $\mathbf{n}$. 
We introduce a vanishing mobility and shear stress when the GB approaches the free surface to mimic its expected behavior there. 
For further details of the model and the numerical method (finite difference with semi-implicit time integration) employed to simulate its evolution dictated by Eq.~\eqref{eq:pf-dynamics}, we refer to \cite{sal2022disconnection}.

Numerical simulations of capillarity-driven grain growth (mean curvature flow) for a microstructure comprising many grains are also considered to mimic the known effect of surface pinning (mimicking thermal grooving), and are then compared with scenarios that include elastic effects and the related impact of a free surface. To this goal, we consider a standard multi-phase field model \cite{Steinbach1999,Steinbach_2009}. A set of $N$ order parameter fields $ \varphi_1(\mathbf{r}), \varphi_2(\mathbf{r}), \ldots,\varphi_{N}(\mathbf{r})$ with $\varphi_i(\mathbf{r})$ that is 1 inside the $i$-th grain and 0 outside, with a smooth transition in between.
The evolution of the order parameter field $\varphi_i$, assuming isotropic mobility and interface energy, reads
\begin{equation}\label{eq:pf-dynamic2}
\begin{split}
    \frac{\partial \varphi_i}{\partial t} &= \sum_j^{N} \left[\varphi_j \nabla^2\varphi_i - \varphi_i \nabla^2\varphi_j - \frac{\pi^2}{2\eta^2}\left(\varphi_j - \varphi_i\right)  \right].
\end{split}
\end{equation}
$\eta$ is a parameter that relates to the GB thickness. In the simulations performed in this work, we set $\eta=10$. Simulations are initialized from a microstructure containing 1000 grains in a square domain, which are first relaxed using Eq.~\eqref{eq:pf-dynamic2} under periodic boundary conditions (PBCs). Surface pinning is then modeled by freezing the microstructure along the boundaries normal to the $y$-direction (Dirichlet boundary conditions), while retaining PF dynamics along boundaries normal to the $x$-direction. Additional numerical details are provided in~Ref. \cite{qiu2025grain}.

\FloatBarrier

\section{Results}

\subsection{Gradient microstructural evolution}
\label{Gradient}
\begin{figure}[htp!]
    \centering
    \includegraphics[width=0.95\textwidth]{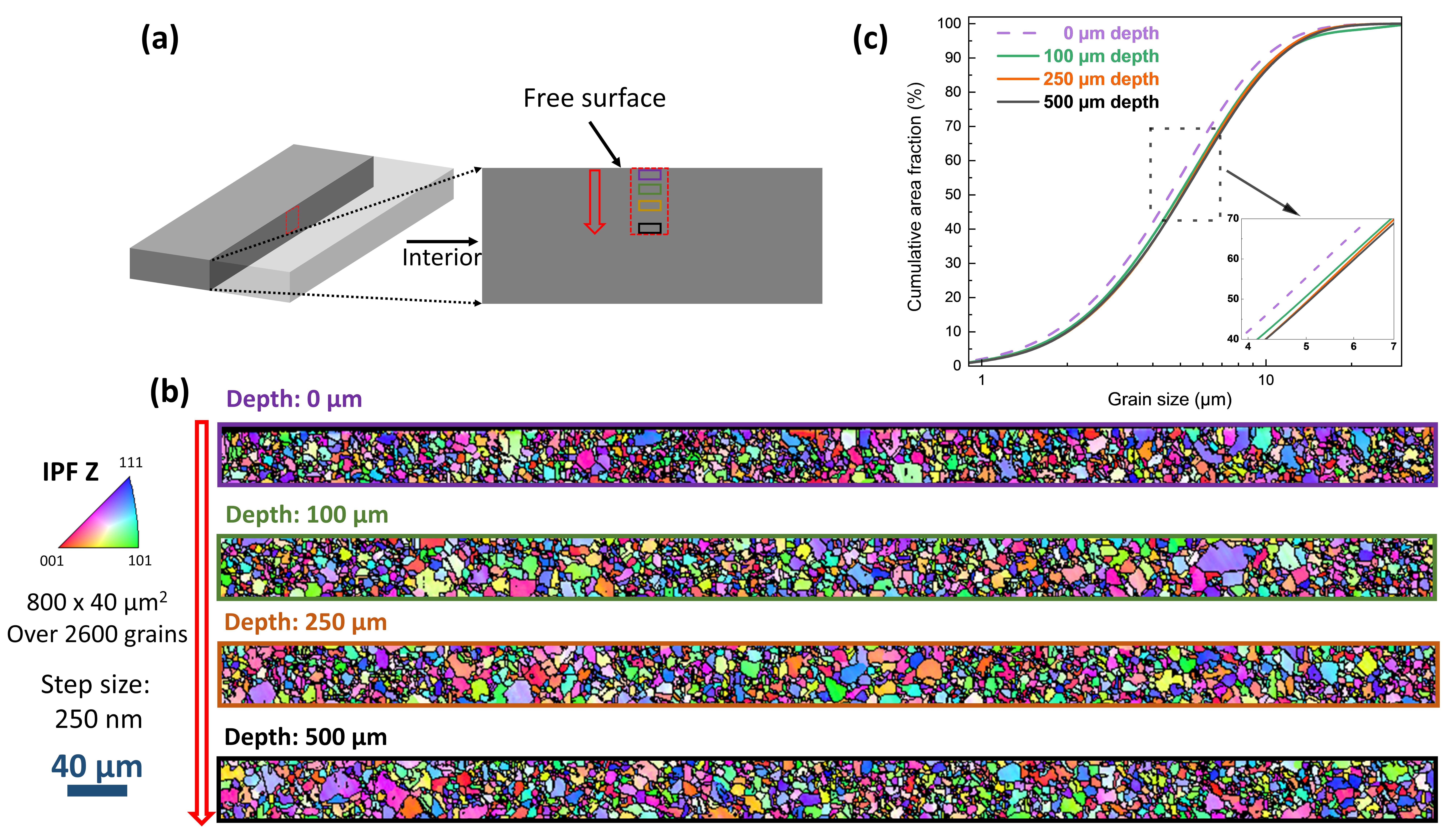}
    \caption{EBSD orientation maps and corresponding grain size distribution curves obtained on specimen set \textit{i} after annealing at 400 \degree C  for 1 h. (a) Schematic depicting the analyzed sample regions; (b) Detailed EBSD orientation maps from the regions in (a); (c) Cumulative grain size distribution curves based on the data set displayed in (b); Considering that about 2600 grains were analyzed for each depth, there is a clear tendency for larger grains in the specimen interior ($>$ 100 $\mu$m depth). The inset further highlights this trend in a clearer manner. Moreover, retardation of grain growth is not restricted to the very specimen surface but prevails for several layers of grains. The color code of orientation maps corresponds to IPF in the direction perpendicular to the sample surface.}
    \label{figure4}
\end{figure}

Fig.~\ref{figure4} displays the EBSD orientation maps (with color code corresponding to inverse pole figure (IPF) in the direction perpendicular to the sample surface) obtained on sample set \textit{i} (see Fig.~\ref{figure1} for details). The maps were acquired at different depths from the surface. The sample was annealed at 400 \degree C for 1~h. Already from the EBSD orientation maps (Fig.~\ref{figure4}(b)), a distinct microstructural evolution, with a higher probability of smaller grains near the free surface, can be observed. This retardation in growth kinetics is not only observed at the very surface, where drag from GB grooving might play a role, but throughout the first analyzed layer with a thickness of 40 $\mu$m. This trend is also captured in the corresponding cumulative area-fraction grain-size distribution (GSD) curves (Fig.~\ref{figure4}(c)), which are based on more than 1500 grains over an area of $800 \times 40$ $\mu$m$^{2}$, respectively. Grains in the first 40 $\mu$m from the specimen surface had an overall reduced grain size compared to the regions a few 100 $\mu$m away from the specimen surface, where no further grain size changes were observed with increasing depth. The retardation of grain growth within the first 40 $\mu$m thick layer, differed from the rest (bulk) of the specimen. In brief, as discussed further below, free surfaces are found to affect grain size beyond capillarity effects.

\FloatBarrier

\begin{figure}[htp!]
    \centering
    \includegraphics[width=0.95\textwidth]{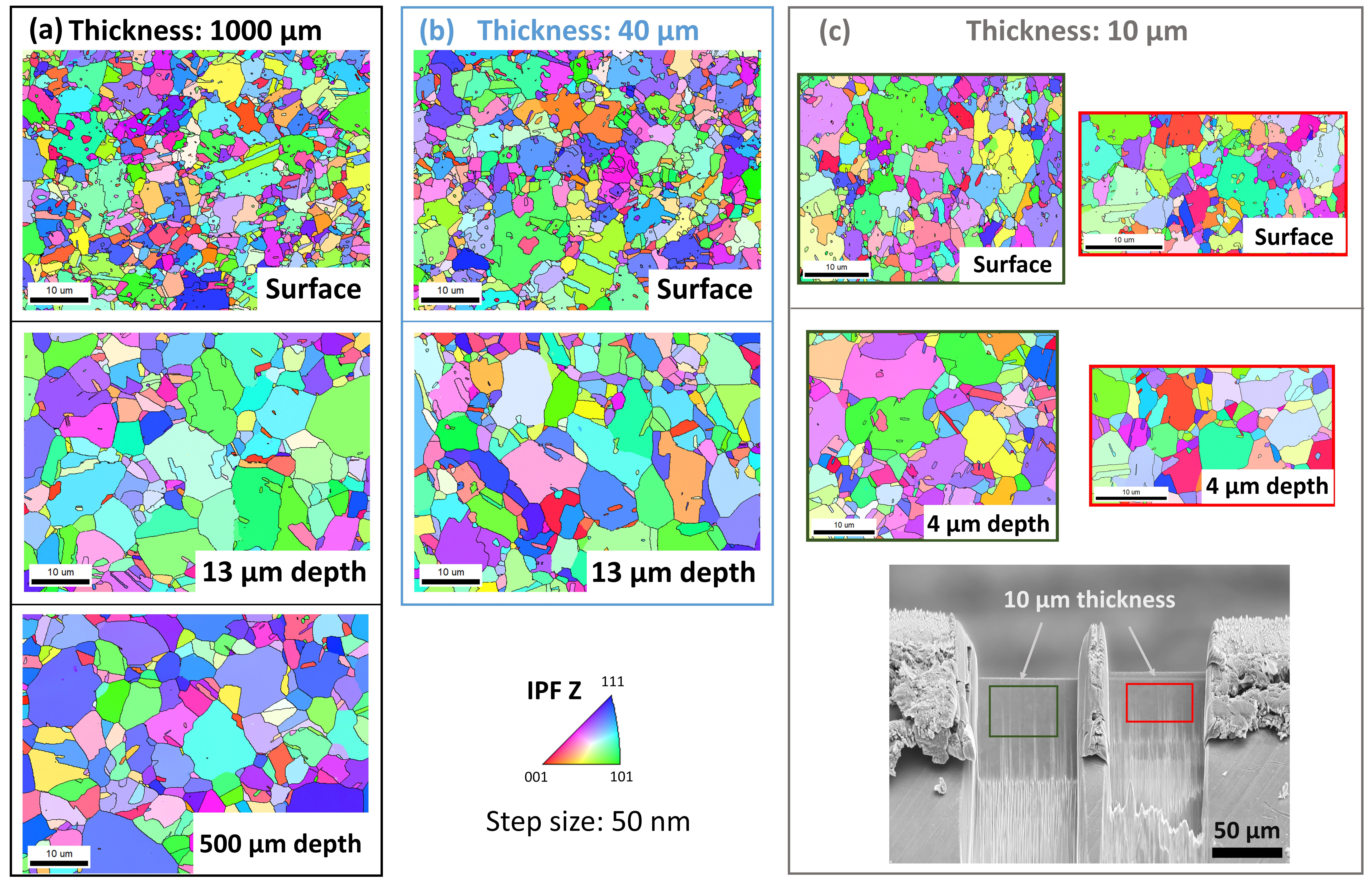}
    \caption{Representative EBSD orientation maps taken at different depths from the surface of a 1 mm thick sample (a); of a 40 $\mu$m thick specimen (b) and of two ultra-thin lamellae taken at the very surface and after removing additional 4 $\mu$m thickness by FIB polishing (c). The color code of orientation maps corresponds to IPF in the direction perpendicular to the sample surface.}
    \label{figure5}
\end{figure}

Expanding further on this finding, we evaluated the microstructural evolution of specimen set (ii) at different depths in a sequential manner, as shown in the schematic in Fig.~\ref{figure2}. In Fig.~\ref{figure5}(a), representative EBSD orientation maps after an annealing treatment at 400 \degree C for a 1000 $\mu$m thick sample taken at the very surface (0 $\mu$m depth), at 13 $\mu$m and 500 $\mu$m depth, are displayed, respectively. Directly at the surface, many small island-like grains seem to remain embedded within the grown, larger grains. Presumably, this feature is a result of GB drag by thermal grooving, as, already at 13 $\mu$m depth from the surface, the island-like grains were widely, but not completely, absent, and the grain size was larger than at the surface. These island-like grains were absent at 500 $\mu$m depth. 

Similar microstructural differences as a function of increasing depth from the surface were obtained for the 40 $\mu$m thick sample, as shown in Fig.~\ref{figure5}(b). Grains are significantly larger in the interior of the specimen, while island grains can still be found. Backscattered electron images taken on a cross section of one of the 40 $\mu$m thick specimens confirmed the thickness and indicated that about five to ten grains spanned across the thickness, see Fig.~\ref{figureS1}. Eventually, a qualitatively similar behavior was observed for the thinnest specimens (i.e., the FIB milled and only 10 $\mu$m thick lamellae), see Fig.~\ref{figure5}(c).

\FloatBarrier

\begin{figure}[htp!]
    \centering
    \includegraphics[width=0.95\textwidth]{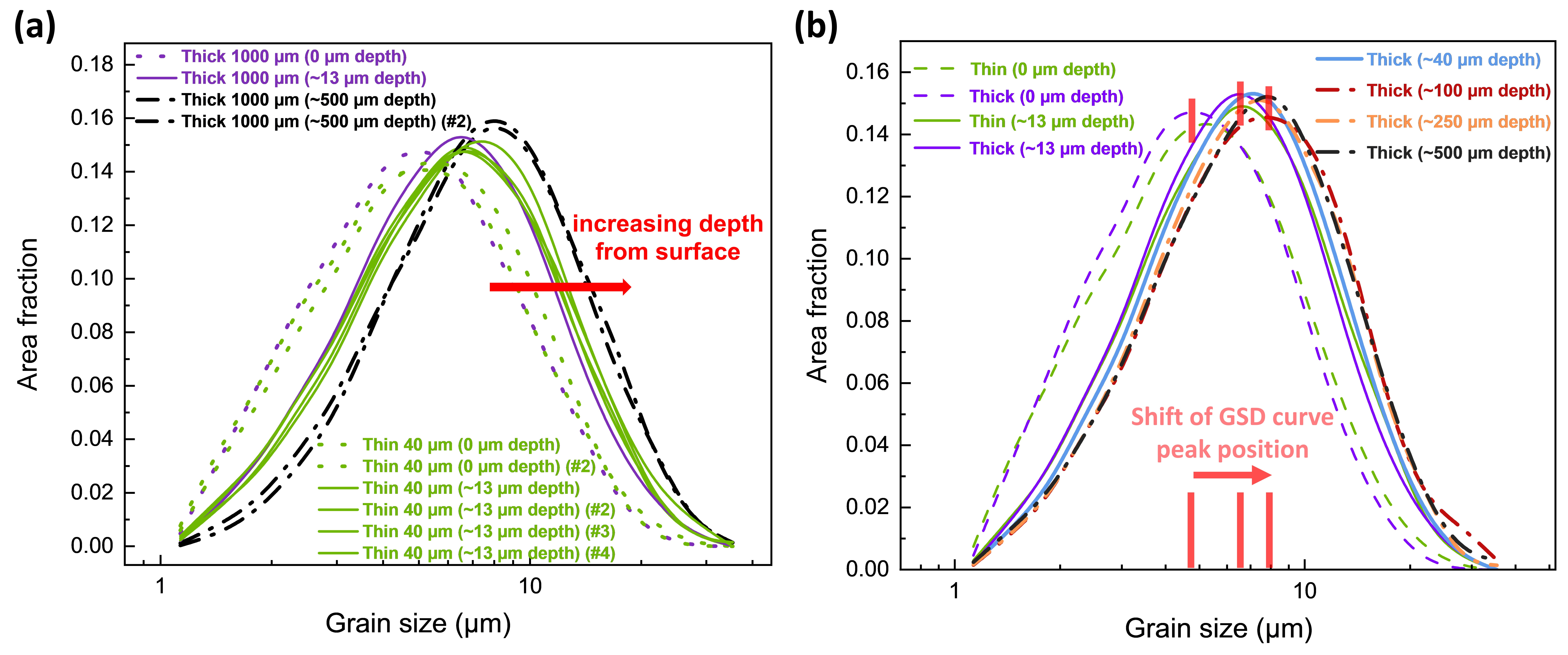}
    \caption{(a) Representative grain size distribution curves of various 1 mm and 40 $\mu$m thick samples obtained at different representative depths from the surface. Each distribution contains grain size data from EBSD scans covering about 30,000 grains. \#2, \#3, etc. refers to the individual specimen number, respectively; (b) Plot of all the grain size distributions of one 1 mm thick and one 40 $\mu$m thick specimen at different depths from the surface (compare Fig.~\ref{figure2}). A gradient in grain size evolution (growth kinetics), with increasing grain size towards the interior, can be clearly noticed. The short red lines indicate the peak position of the GSD curves at 0 $\mu$m, 13$\mu$m and 500 $\mu$m depth from the surface.}
    \label{figure6}
\end{figure}

Similar to Fig.~\ref{figure4}, the trends of larger grains in the specimen interior became clearly visible when a sufficient number of grains were analyzed. Grain size distributions based on large area EBSD scans (872 $\mu$m field of view with 500 nm step size, see Supplementary Fig.~\ref{figureS2}), show that this is indeed the case, as reported in Fig.~\ref{figure6} (a). A clear difference between the grain size distributions of the thick samples (1 mm thickness) at 13 $\mu$m and 500 $\mu$m depth emerged. Moreover, there is no difference in grain size between the thin (40 $\mu$m thickness) and thick samples (1 mm thickness) directly at the surface, nor in 13 $\mu$m depth. However, grain sizes in the specimen interior (at 500 $\mu$m depth) of the 1 mm thick sample are considerably larger than those in the center of the thin specimens (at 13 $\mu$m depth) having 40 $\mu$m thickness. By observing these results over different samples in Fig.~\ref{figure6}(a), we conclude that the observed trend of increasing grain size toward the specimen interior is not a random variation over measurements but is related to differences in the growth kinetics. Data obtained from additional samples, from several different regions within a sample, and from different depths below the surface confirm the aforementioned findings and are presented in the supplement (see Fig.~\ref{figureS3}). Individual scans from a single sample yielded highly consistent grain size measurements, compare Fig.~\ref{figureS3}.

Plotting all data at different depths obtained on two selected different specimens in Fig.~\ref{figure6}(b) clearly shows the discussed gradient in microstructural evolution. While there is little difference in grain size distribution at the surface between the thick and thin specimens, there is already a clear increase in size for data obtained at 13 $\mu$m depth. In the case of the 1 mm thick sample, a slight further increase in grain size can be noticed when analyzing the microstructure at 40 $\mu$m depth. This trend continues at even larger depths, but the increase in grain size tends to  saturate for depths $> 100$ $\mu$m from the surface.

\FloatBarrier

\begin{figure}[htp!]
    \centering
    \includegraphics[width=0.95\textwidth]{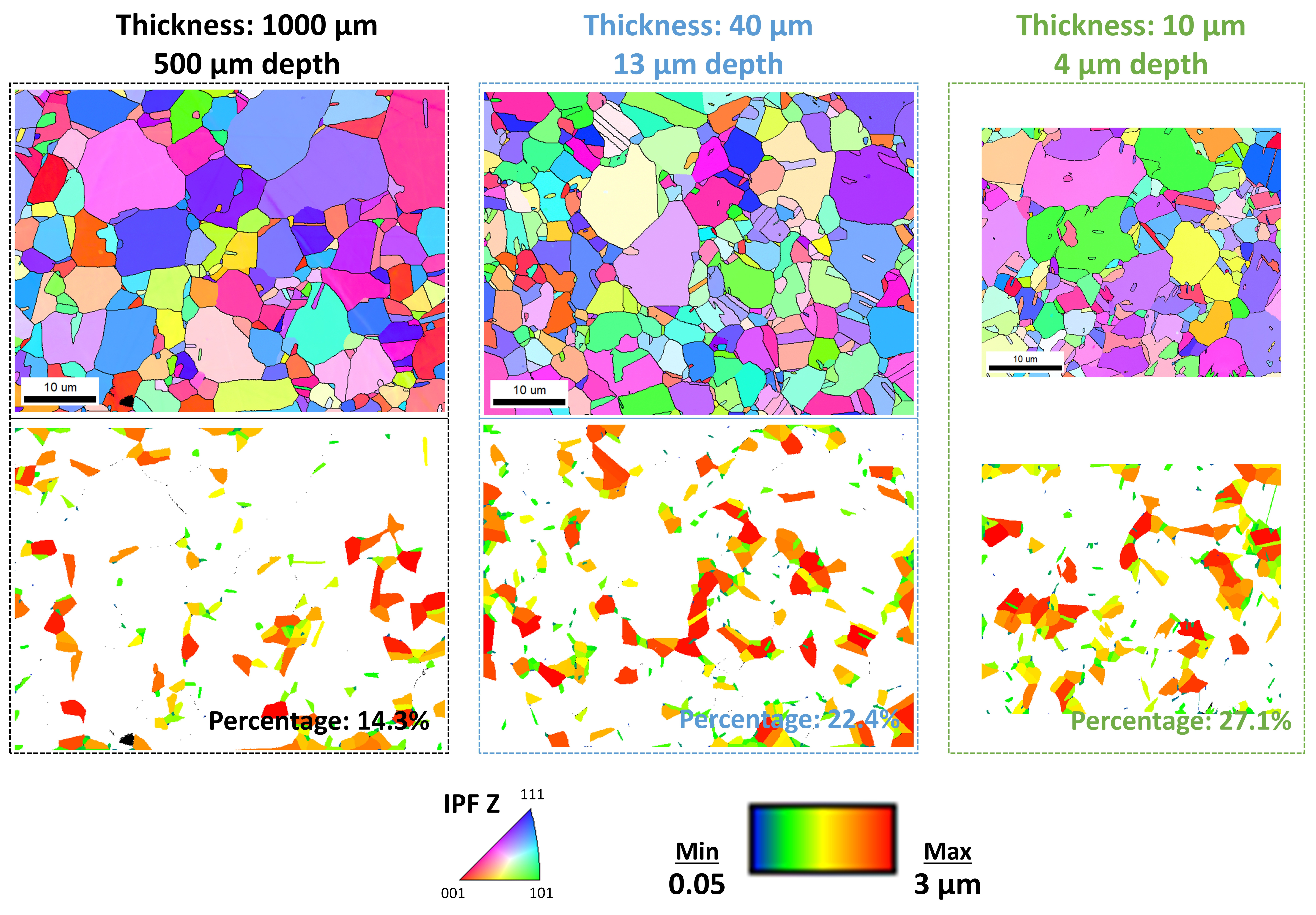}
    \caption{Comparison of the fraction of small grains (\textless 3 $\mu$m size, with $\Sigma3$ twin boundaries included) surviving the heat treatment in the specimen center (at approximately half thickness) for different sample thicknesses. Grain growth is clearly subdued as specimen size decreases. The scale bars for the two maps within a particular dotted-line frame are the same. The color code of EBSD orientation maps corresponds to IPF in the direction perpendicular to the sample surface.}
    \label{figure7}
\end{figure}

Accordingly, the specimen size can have a profound effect on the growth kinetics, if microstructures are analyzed at half-thickness; see Fig.~\ref{figure7}. EBSD orientation maps obtained at approximately half-thickness of the respective samples (grooving effects safely excluded) are compared across the three specimen thicknesses studied (1000 $\mu$m, 40 $\mu$m, and 10 $\mu$m). From Fig. \ref{figure7} it is evident that the fraction of small grains ($\Sigma3$ twin boundaries included in grain-size statistics) in the specimen interior (grain size \textless 3 $\mu$m) decreases as the specimen thickness increases. Note that this trend does not change when the thin, lamellar annealing twins are excluded from the grain size statistics (Supplementary Fig. \ref{figureS4}). In fact, for the 10 $\mu$m thick specimen, the fraction of small grains is almost 90 \% larger compared to the 1 mm thick specimen. While this pronounced difference might be largely attributed to drag effects from thermal grooving for this very thin specimen, a similar tendency can still be observed for the 40 $\mu$m thick specimen. Even for this sample, the fraction of small grains surviving the annealing treatment in the specimen center was about 50\% larger compared to the 1 mm thick specimen (at 500 $\mu$m depth).

\FloatBarrier

\subsection{Impact of a free surface on grain boundary migration}

\begin{figure}[htp!]
    \centering
\includegraphics[width=0.95\textwidth]{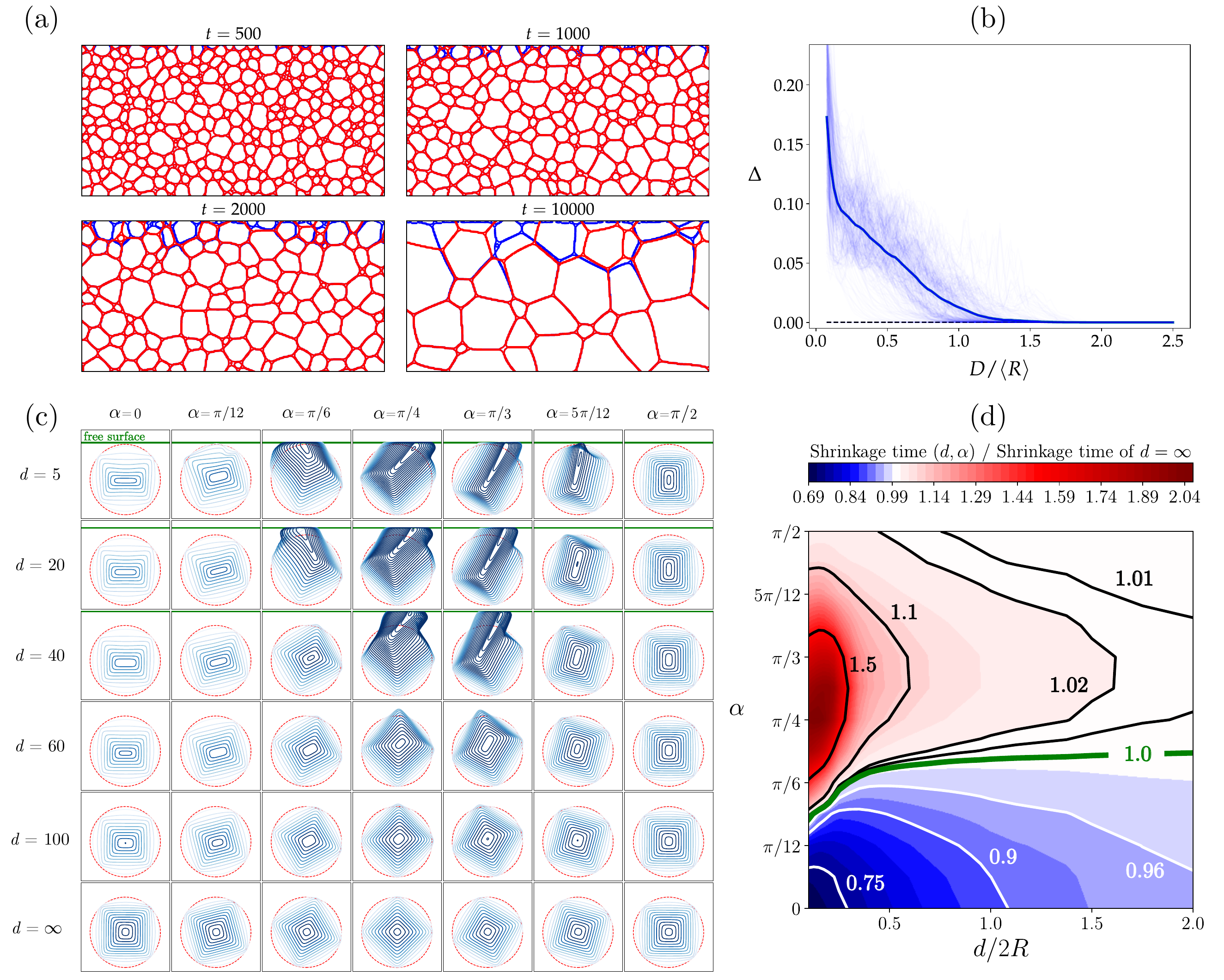}
    \caption{(a) Illustration of grain growth by mean-curvature flow with (blue) and without (red) surface pinning. 
(b) ${L}^2$ norm of the difference between the fields corresponding to the networks with and without pinning, as shown in panel (a), plotted as a function of the distance from the surface $D$, normalized by the average grain size $\langle R \rangle$. The solid blue line denotes the average over different times (every $\Delta t = 250$) and 10 independent simulations; individual realizations are shown in the background with low opacity. (c) Dynamics of an initially circular GB (red dashed line) in the presence of a free surface (green solid line). Light blue lines represent the shrinkage of this circular GB. The morphological evolution by varying distance from the surface ($d$) and orientation with respect to the free surface ($\alpha$) is shown. Color scale from (d) indicates the ratio between the annihilation time for given $d$ and $\alpha$ and the limit for $d\rightarrow \infty$. Accordingly, regions highlighted in red correspond to scenarios slowing down the annihilation.}
    \label{figure8}
\end{figure}

The systems discussed in the previous sections differ from bulk polycrystals in that they have a free surface. For GBs terminating at such a surface, thermal grooving occurs, which is expected to pin GB migration \cite{mullins1958effect,frost1990simulation,barmak2006grain,verma2021grain}. The limit of strong surface pinning is illustrated in Fig.~\ref{figure8}(a), which shows multi-PF simulations of Eq.~\eqref{eq:pf-dynamic2} reproducing known phenomenology as reference (see Section~\ref{sec:numerical-simulations} for details). The GB network is reported at representative stages for simulations with (blue) and without (red) surface pinning. As expected, the impact of pinning increases with time, while it remains confined to a depth comparable to the average grain size. This behavior is quantified in Fig.~\ref{figure8}(b), which reports the ${L}^2$ norm ($\Delta$) of the difference between the networks shown in Fig.~\ref{figure8}(a), averaged every $\Delta t = 250$ over ten independent simulations, as a function of the depth normalized by the average grain size $\langle R \rangle$. The averaged $\Delta$ approaches zero for $\langle R \rangle \sim 1.25$. Therefore, drag effects from thermal grooving alone cannot account for the experimentally observed depth-dependent microstructural evolution, which extends down to depths of almost 40 $\mu$m, i.e., at least five grain layers, while only for depths larger than 100 $\mu$m a uniform grain size distribution was observed.

 The presence of a free surface also affects the elastic field of disconnections (or dislocations) as shown in Sect.~\ref{sec:continuum-modeling}. Leveraging the model outlined therein, we inspect the impact of a free surface on the morphological evolution of an idealized grain within the matrix (crystal). We consider a circular GB (so a rotated grain in a matrix) in the presence of a straight free surface. We simulate its evolution by varying the distance ($d$) from the free surface, and the relative orientation of the Burgers vector of the disconnection with respect to the surface normal (quantified by the angle $\alpha$), mimicking a randomly oriented grain in a matrix (see also Fig.~\ref{figure3} for illustration of the corresponding geometrical parametrization). Fig~\ref{figure8}(c) shows the morphological evolution for representative settings, while Fig~\ref{figure8}(d) illustrates the ratio of the disappearance time of the evolving grain in the systems with the free surface compared to bulk for varying $d$ and $\alpha$. As expected, for very large distances (namely $d\rightarrow \infty$, effectively in the bulk), the impact of the free surface is negligible, and the morphological evolution is self-similar, resulting just in the rotation of the facets developing due to the shear coupling \cite{han2022disconnection,sal2022disconnection,qiu2023interface}. When approaching the free surface, but at relative distances significantly larger than in Figs ~\ref{figure8}(a,b), the symmetry breaking induced by the additional condition on the elastic field becomes effective.  Interestingly, for this isolated grain, a non-trivial scenario involving both acceleration and deceleration of grain growth emerges. For $\alpha=0$ and $\alpha=\pi/2$, the evolution is slightly faster and slightly slower than in bulk, respectively, although minor morphological changes occur with respect to the bulk limit. For intermediate orientations, the effect is more pronounced: driven by the shear coupling with an increased stress and significantly opposite sign (see also Fig.~\ref{figure3}(c) for $\alpha=\pi/4$), the free surface attracts the GB, eventually leading to a slower dynamical process. Note that in Fig~\ref{figure8}(d) a slower dynamic than in bulk is observed across a significantly larger number of configurations (as can be seen by comparing the areas where the relative shrinkage time is greater or less than one). Moreover, the shrinkage of the grain takes double the time in the slowest case ($\alpha=\pi/4$), compared to a reduction time of $30\%$ for the fastest case. These values might change slightly when surface grooving (neglected in this setting) is considered, but this would not affect the trends discussed, which largely refer to stages for which the GBs do not meet the free surface. We also remark that the relative impact of shear coupling vs capillarity is expected to increase with the grain size \cite{qiu2026shear}, so the effect is not expected to vanish during grain growth, except when entering different regimes (e.g., for plastic relaxation of the grains). 
 
In real microstructures, the impact of shear coupling is more complex. Even under a single-disconnection-mode assumption, grains experience different shear-coupling behaviors at each delimiting grain boundary \cite{qiu2025grain}. A fully self-consistent description that accounts for realistic free-surface morphologies (including surface grooving) and the associated elastic relaxation remains beyond current modeling and computational capabilities. Still, the simplified analysis presented here already clarifies a key mechanism governing variations in internal stress induced by free surfaces. For isolated grains, grain growth is expected to be slower near free surfaces. More importantly, the depth of the region affected by shear-coupled grain boundary migration extends well beyond that associated with capillarity (grooving). This can be attributed to the fact that the intrinsically long-range (dislocation-like) elastic field of disconnections mediating shear coupling becomes effectively short-ranged in the presence of a free surface, with a decay that scales with the distance from the surface \cite{head1953edge}. Consequently, the free surface influences the internal stress, whether or not grain boundaries terminate at it. Its impact on the dynamics progressively diminishes with increasing depth, as observed in simulations, but retains a longer range than capillarity effects. We also obtained that grain growth might exhibit a broader size distribution than in bulk, since the orientation of the shear stress coupled to grain boundary migration may either enhance or retard grain shrinkage, although confirming this experimentally would require highly accurate statistics.

\FloatBarrier

\section{Discussion}

By investigating grain growth kinetics near free surfaces in polycrystalline nickel, we observed a clear gradient in grain size evolution, with smaller average grain sizes closer to the free surface. Cross-sectional and in-plane analyses consistently showed that this gradient extended to depths of more than five grain layers. This indicates that the free-surface effect is not restricted to the immediate vicinity of the surface but influences grain growth over a substantially larger distance.

While surface effects on grain growth are rarely considered in case of bulk, polycrystalline specimens, retarding effects on grain growth by free surfaces have been frequently observed for thin-film materials. In the latter case, grain growth is often observed to quickly stagnate once the grain size reached a critical fraction of the film thickness \cite{an2009kinetic, simoes2010situ, palmer1987grain, nichols1993situ}. This growth stagnation, typically occurring when the grain size reached two to three times the film thickness \cite{verma2021grain}, has been widely attributed to the formation of thermal grooves at the intersection of GBs with the free surface \cite{verma2023effect}. However, considering the depth of the surface grooves, typically in the range of 50-100 nm as measured by atomic force microscopy \cite{ding2021situ}, a pronounced influence of these grooves on grain growth in larger depths from the surface may not be expected.

Our simulations suggest that the effect of thermal grooves on GB migration rapidly diminishes with increasing depth (relative to grain size) and can already be considered negligible after the first layer of grains, compare Figs. \ref{figure8}(a,b). Note, that the conditions applied in this simulation can be considered more restrictive compared to real scenarios, as the microstructure was frozen in a small region at the free surface. Surface grooves may thus explain growth stagnation in thin-film materials with only one or a few grains across the film thickness. This, however, does not apply to the specimens studied here, where surface-grooving-induced GB drag alone cannot explain the observed gradient microstructure evolving over much larger depths. Different origins have thus to be taken into account. It is worth noting that, also for thin film specimens, thermal grooves may not be the only origin of growth stagnation, as stagnation has also been observed when protective layers have been applied onto the film \cite{barmak2006grain,rabkin2020grain}. 

Recent analysis indicates that stress buildup from shear-coupled grain growth in the films could be the general cause of growth stagnation \cite{rabkin2020grain}. Since shear-coupled migration upon grain growth was also observed for nickel at similar temperatures to those applied in this work \cite{tang2026atomic}, the associated intrinsic stresses, which affect growth kinetics \cite{qiu2025grain}, could vary close to the free surfaces. Our simulations suggest that this is indeed the case. Similar to lattice dislocations, the stress fields of disconnections are significantly affected when they are close to the free surface compared to the specimen interior. This change further depends on the orientation of the Burgers vector relative to the free surface, i.e., on the direction along which shear deformation occurs, compare Figs. \ref{figure3} and \ref{figure8}. This change in the internal stress could lead to an acceleration or retardation of grain growth kinetics. For an isolated grain, the number of cases that lead to acceleration is rather limited, see Fig. \ref{figure8}(d). In any case, the effects of this stress relaxation or modification on grain growth near free surfaces reach much deeper into the bulk than in the case of thermal grooving. Larger affected depths, as well as the experimental result that not all surface grains grow much more slowly (Fig. \ref{figure3} cross section), are well consistent with these simulations (Fig. \ref{figure8}(d)). Recalling the minimal nature of the considered theoretical framework, we note that specific morphologies of evolving grains can in principle be accounted for by considering an extended set of reference interfaces ($k$), which implies considering an extended set of disconnections \cite{qiu2023interface}, and multi-disconnection modes \cite{qiu2024disconnection}), although the qualitative behavior of shear-coupled migration as intended here is well described by the assumption of the reported theoretical analysis for both single grains and microstructures \cite{qiu2025grain,qiu2026shear}. Instead, an extension of the framework for studying analogous settings in 3D has not yet been proposed. This could be achieved by extending the current model, provided that a proper description of disconnection lines and their evolution can be devised, or by directly encoding shear deformation through eigenstrain formulations.

We therefore conclude that thermal grooving alone cannot account for the observed gradient microstructures, pointing instead to variations in internal stresses during grain growth in the presence of free surfaces as a key factor, particularly when the number of grains across the sample becomes limited (at least five layers in the nickel samples studied here), yet still above the ultrathin (semi-2D) limit.

State of the art 3D grain growth studies (e.g., by 3D-XRD technique \cite{schmidt2008direct}), which typically deal with rather thin platelets or needle-shaped specimens with minimum dimensions on the order of several 100 $\mu$m to 1 mm, fall in case of typical grain sizes (several 10 $\mu$m – less than 10 grain layers across thickness) already into the range where for nickel pronounced effects from the free surface on the microstructural evolution could be observed. This indicates that such data, as well as their comparison with microstructural evolution or kinetics observed in bulk 2D sections, must be carefully interpreted. Recent advances in the field towards thicker specimens and larger analyzed volumes using pink X-ray beams are thus highly appreciated \cite{yildirim20253d}.

\section{Summary and outlook}
In the present work, we report on an intrinsic gradient microstructure that developed upon grain growth of bulk, polycrystalline nickel in the vicinity of the free surface. For depths up to about 40 $\mu$m from the free surface, corresponding to at least five layers of grains, grain growth was clearly retarded compared to the bulk. Similarly, specimens with thicknesses less than 40 $\mu$m also showed slower growth kinetics under the same heat treatment than bulk specimens when analyzed at mid-sample height. Phase-field simulations indicate that thermal grooving at the free surfaces could only explain sluggish growth for the first grain layer at maximum. In contrast, the modification of internal stresses from shear-coupled GB migration induced by free surfaces could affect grain growth at much larger depths. For isolated grains, the orientation of the shear deformation can induce either an acceleration or a retardation of growth kinetics, with the majority of cases corresponding to the latter case. A more quantitative description will become attainable once a self-consistent framework for polycrystalline grain growth in the presence of free surfaces is established. Tailoring this effect would also require additional experimental settings involving more than one free surface. Future experiments and simulation work are thus necessary to understand in detail how free surfaces affect grain growth in polycrystals, building on the evidence reported here. This may be relevant not only to specimens with reduced dimensions but also to those in which yield or fatigue could be largely affected by near-surface grains.

\FloatBarrier

\section*{Acknowledgment}

OR gratefully acknowledges financial support by the Austrian Science Fund via project PIN4577425 and the Austrian Federal Ministry for Digital and Economic Affairs, the National Foundation for Research, Technology and Development and the Christian Doppler Forschungsgesellschaft (CDG).\\
MS acknowledges funding by the Deutsche Forschungsgemeinschaft (DFG, German Research Foundation) Project No.~447241406, and the computing provided by the NHR Center of TU Dresden.\\

\section*{CRediT authorship contribution statement}
\textbf{Jing Tang}: Conceptualization, Methodology, Investigation, Visualization, Writing --- original draft.
\textbf{Runlu Yan}: Investigation.
\textbf{Dong-lan Zhang}: Writing --- reviewing and editing.
\textbf{Ronald Schnitzer}: Resources, Writing --- reviewing and editing.
\textbf{Lorenz Romaner}: Writing --- reviewing and editing.
\textbf{Marlene Kapp}: Conceptualization, Resources, Writing --- reviewing and editing.
\textbf{Marco Salvalaglio}: Conceptualization, Funding, Methodology, Investigation, Visualization, Writing --- original draft, Writing --- reviewing and editing. 
\textbf{Oliver Renk}: Conceptualization, Funding, Methodology, Investigation, Writing --- original draft, Writing --- reviewing and editing, Supervision.

\section*{Disclosure statement}

No potential conflict of interest was reported by the authors.

\section*{Data availability}
The raw data sets required to reproduce the findings of the present work are available upon reasonable request.

\section{Appendices}

\noindent\textbf{Supplementary information}\medskip

\renewcommand{\thefigure}{S1}
\begin{figure}[htp!]
    \centering
    \includegraphics[width=0.95\textwidth]{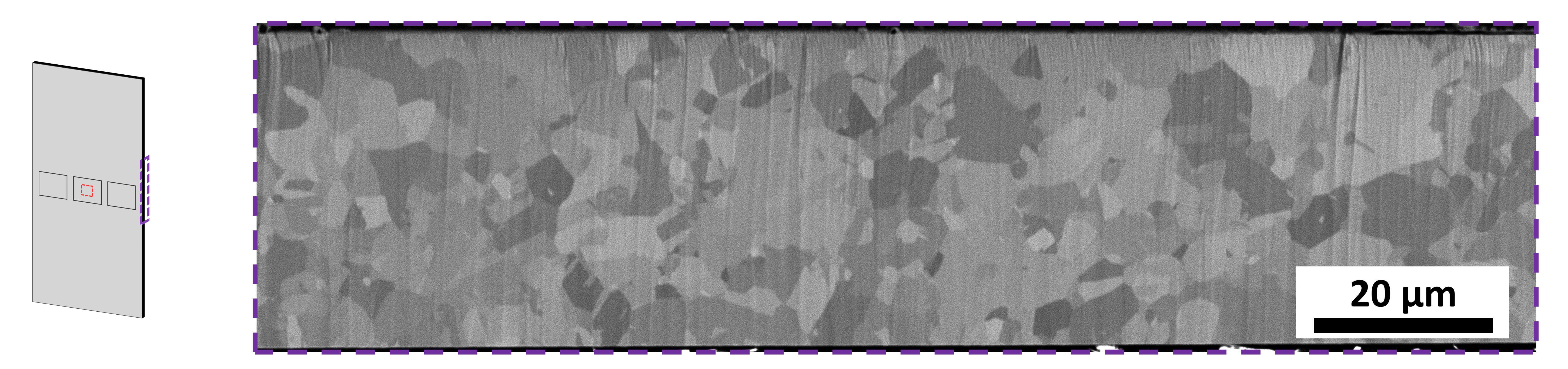}
    \caption{Backscattered electron image of the cross section of an annealed 40 $\mu$m thick sample, confirming its polycrystalline character, with five to ten grains across the thickness.}
    \label{figureS1}
\end{figure}

\renewcommand{\thefigure}{S2}
\begin{figure}[htp!]
    \centering
    \includegraphics[width=0.95\textwidth]{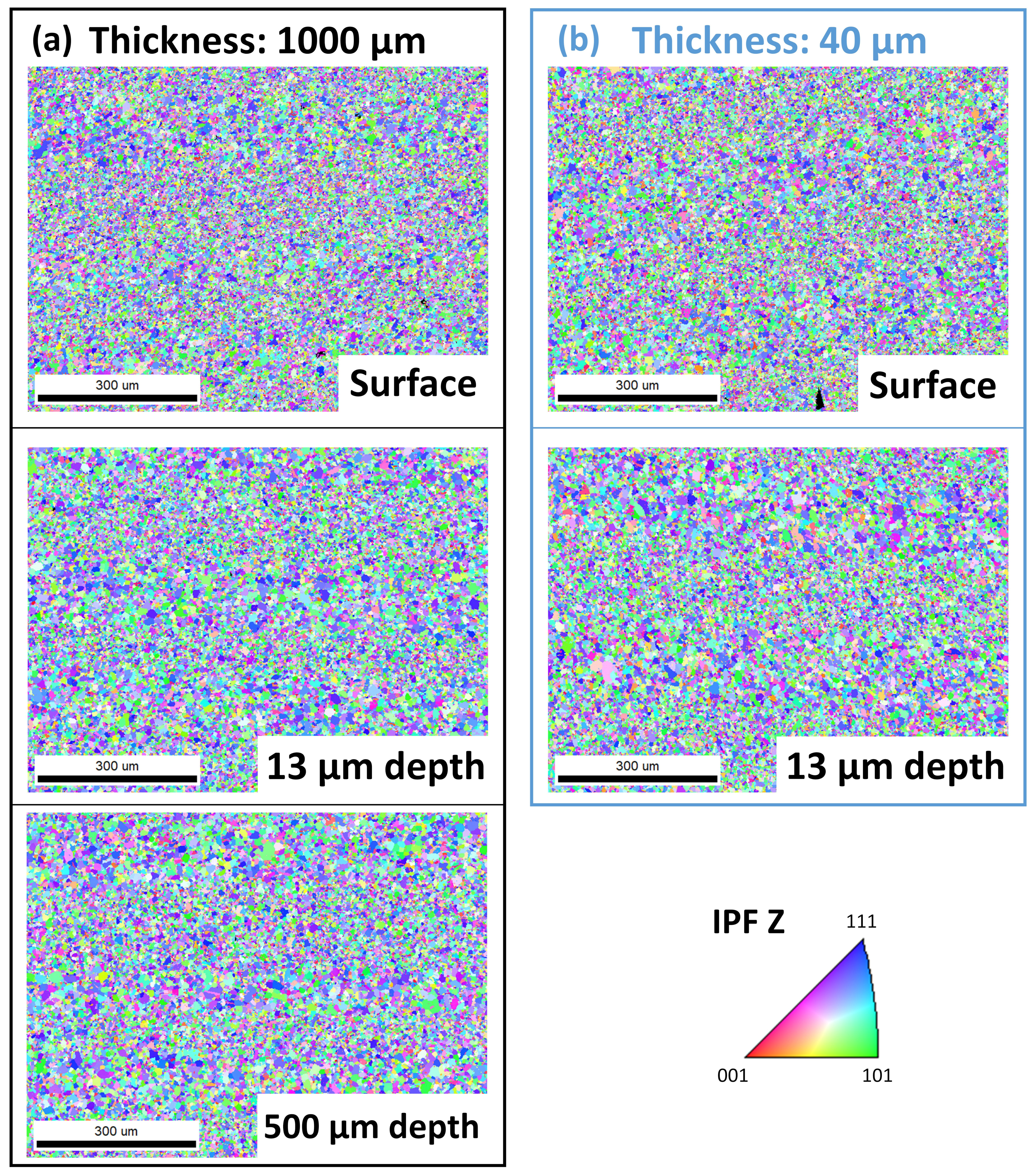}
    \caption{(a) Representative large area EBSD orientation maps taken at different depths with respect to the free surface of a 1 mm thick sample; (b) Representative large area EBSD orientation maps taken at different depths with respect to the free surface of a 40 $\mu$m thick sample; The color code of orientation maps corresponds to IPF in the direction perpendicular to the sample surface. The step size was set to 500 nm in all scans.}
    \label{figureS2}
\end{figure}
\renewcommand{\thefigure}{S2}

\renewcommand{\thefigure}{S3}
\begin{figure}[htp!]
    \centering
    \includegraphics[width=0.95\textwidth]{Figures/S_Figures/Figure_S3.jpg}
    \caption{Grain size distribution (GSD) curves for several 1 mm thick and 40 $\mu$m thick specimens, taken from different areas and depths from the surface, respectively. \#1 and \#2 label two different samples. (a) GSD curves based on large-area EBSD scans (about 30000 grains per scan) of two 1000 $\mu$m thick specimens (3 areas respectively) taken at 500 $\mu$m depth from the surface; (b) GSD curves based on large-area EBSD scans (about 50000 grains per scan) of two 40 $\mu$m thick samples obtained directly on the surface; (c) GSD curves based on large-area EBSD scans (about 30000 grains per scan) of two 40 $\mu$m thick samples at 13 $\mu$m depth from the surface. The distributions show marginal scatter when considering a particular depth from the surface, but significant differences between the distributions at different depths clearly appear. Quantitatively, taking sample \#1 in Fig. S3(a) as an example, the three scans yield average grain sizes of 8.31 $\mu$m, 8.78 $\mu$m, and 8.55 $\mu$m, with an overall average of 8.55 $\mu$m, and standard deviations of 5.01 $\mu$m, 5.08 $\mu$m, and 5.02 $\mu$m, respectively.}
    \label{figureS3}
\end{figure}

\renewcommand{\thefigure}{S4}
\begin{figure}[htp!]
    \centering
    \includegraphics[width=0.95\textwidth]{Figures/S_Figures/Figure_S4.jpg}
    \caption{Comparison of the fraction of small grains (\textless 3 $\mu$m) obtained using different lower size limits (50 nm and 2 $\mu$m, respectively), to largely exclude the very fine lamellar twin boundaries. As highlighted by the circles, some small twin-related grains were excluded from the statistics when the lower size limit was set to 2 $\mu$m. Nevertheless, the difference between the two datasets remains pronounced, with approximately 50\% more small grains observed in the thinner sample.}
    \label{figureS4}
\end{figure}

\FloatBarrier

\bibliographystyle{Mystyle}
\bibliography{interactnlmsample}

\end{document}